\documentclass[aps,prl,reprint,amsmath,amssymb,superscriptaddress]{revtex4-2}

\usepackage{amsmath}
\usepackage{graphicx}
\usepackage{dcolumn}
\usepackage{bm}
\usepackage{multirow}
\usepackage{subfigure}
\usepackage{booktabs}
\usepackage{hhline}  
\usepackage{float}
\usepackage{braket}
\usepackage[colorlinks,linkcolor=blue,anchorcolor=blue,citecolor=blue,urlcolor=blue]{hyperref}
\usepackage[normalem]{ulem}

\begin{document}


\title{Coexistence of Superconductivity and Superionicity in Li$_2$MgH$_{16}$}


\author{Haoran Chen}
\affiliation{International Center for Quantum Materials, Peking University, Beijing 100871, China}

\author{Junren Shi}
\email{junrenshi@pku.edu.cn}
\affiliation{International Center for Quantum Materials, Peking University, Beijing 100871, China}
\affiliation{Collaborative Innovation Center of Quantum Matter, Beijing 100871, China}


\date{\today}
	
\begin{abstract}
	We study superconductivity in the superionic phase of the clathrate hydride Li$_2$MgH$_{16}$, where hydrogen ions diffuse among the lattice formed by lithium and magnesium ions.
	By employing the stochastic path-integral approach, we non-perturbatively take into account the effects of quantum diffusion and anharmonic vibrations.
	Our calculations reveal strong electron-ion coupling ($\lambda(0)=3.7$) and a high superconducting transition temperature ($T_c$) of 277 K under 260 GPa, at which the material is still superionic.
	$T_c$ is significantly suppressed compared with the result $T_c=473$~K obtained from the conventional approach based on the harmonic approximation.
	Our study, based on a first-principles approach applicable to superionic systems, indicates that 
	the superconductivity and superionicity can coexist in Li$_2$MgH$_{16}$.
\end{abstract}


\maketitle


\textit{Introduction.}—
Superconductivity is one of the most fascinating phenomena in condensed matter physics.
The Bardeen-Cooper-Schrieffer (BCS) theory first explains the underlying microscopic mechanism, revealing the important role of ion motion in inducing superconductivity.
The theory predicts that a system tends to exhibit high superconducting transition temperature $T_c$ 
if it has high phonon frequencies and strong electron-phonon coupling (EPC).
In the past decade, the idea has inspired the theoretical predictions and experimental discoveries of hydride high-$T_c$ superconductors under high pressures, with transition temperatures approaching or even exceeding room temperature~\cite{wang_superconductive_2012,duan_pressure-induced_2014,drozdov_conventional_2015,liu_potential_2017,drozdov_superconductivity_2019,kong_superconductivity_2021,li_superconductivity_2022,ma_high-temperature_2022}.
The theoretical investigations usually rely on the harmonic approximations.
However, in these systems, the motions of hydrogen ions can significantly deviate from harmonic vibrations, violating the underlying assumption of conventional approaches.
For example, the crystal structure and $T_c$ of hydride sulfide H$_{3}$S, an experimentally observed high temperature superconductor, cannot be determined correctly if anharmonic and quantum effects of hydrogen ions are neglected~\cite{errea_high-pressure_2015,errea_quantum_2016,chen_stochastic_2022}.
Moreover, quantum effects may cause metallic hydrogen to melt at a temperature below $T_c$, resulting in a superconducting liquid~\cite{liu_superconducting_2020,chen_first-principles_2021,mcmahon_high-temperature_2011,chen_quantum_2013,geng_lattice_2015,jaffe_superconductivity_1981}.

When anharmonic or quantum effects are strong, an intriguing dynamically disordered phase named superionic phase, which is characterized by the diffusion of some ions among the lattice formed by others, may emerge.
The property is widely exploited in solid-state electrolytes~\cite{chandra_superionic_1981,kamaya_lithium_2011,wang_design_2015}.
It is also related with several dynamical phases of ice, and is vital to understanding the phase diagram of water~\cite{benoit_tunnelling_1998,goncharov_raman_1999,goncharov_dynamic_2005,schwegler_melting_2008,millot_nanosecond_2019,ye_dynamic_2021}.
It is natural to ask whether the novel state can coexist with superconductivity.
In recent years, some lithium alloys and clathrate superhydrides are found to possess both properties. 
However, superionicity in these materials occurs at temperatures much higher than the superconducting transition temperatures~\cite{liu_potential_2017,liu_dynamics_2018,wan_as-li_2022,wang_pressure_2022,wan_predicted_2022}.
It is yet to know whether there is a material where superconductivity and superionicity could coexist in a certain temperature regime~\cite{wang_quantum_2023,wang_quantum_2021}.
Li$_2$MgH$_{16}$ is a promising candidate for realizing the coexistence.
According to harmonic calculations, the $Fd\bar{3}m$ phase of the material is predicted to possess strong EPC and a $T_c$ as high as 473~K at 250~GPa~\cite{sun_route_2019}. 
However, later studies based on path-integral molecular dynamics (PIMD) simulations reveal that anharmonicity and quantum effects drive the material to be superionic above 25~K,
with hydrogen ions diffusing between sites~\cite{wang_quantum_2021}. 
This makes the prediction of harmonic calculations unreliable.
A first-principles determination of $T_c$ is hindered by the limitation of conventional computational approaches based on harmonic approximations.

In this Letter, we present a first-principles investigation of superconductivity in the superionic phase of Li$_2$MgH$_{16}$.
To take into account the effects of anharmonic vibrations, quantum fluctuations and ion diffusion,
we apply the stochastic path-integral approach (SPIA), which is a non-perturbative approach without making assumptions of the nature of ion motion~\cite{liu_superconducting_2020,chen_first-principles_2021,chen_stochastic_2022,zhang_nonperturbative_2022}.
We study the electronic structure, electron-ion coupling, and superconductivity in the superionic Li$_2$MgH$_{16}$.
We find that the electronic structure is strongly renormalized by ion diffusion, and the coupling between electrons and ions is strong.
Notably, we observe that the material has a high $T_c=277$~K, at which it is superionic. Therefore, superconductivity and superionicity can coexist in the system.

\textit{Ion motion.}—We begin by examining the ion motion in Li$_2$MgH$_{16}$. To fully take into account the effect of anharmonicity and quantum tunneling, we perform PIMD simulations~\cite{marx_ab_1996,ceriotti_efficient_2010}.
The interaction between ions is described with machine-learning force field (MLFF). Following the original spirit of Refs.~\cite{jinnouchi_--fly_2019,jinnouchi_phase_2019}, the potential field is first trained on the fly in MD simulations, followed by a on-the-fly PIMD training. Detailed descriptions can be found in Supplemental Material~\cite{noauthor_see_nodate}.

\begin{figure}
	\includegraphics[width=8.6cm]{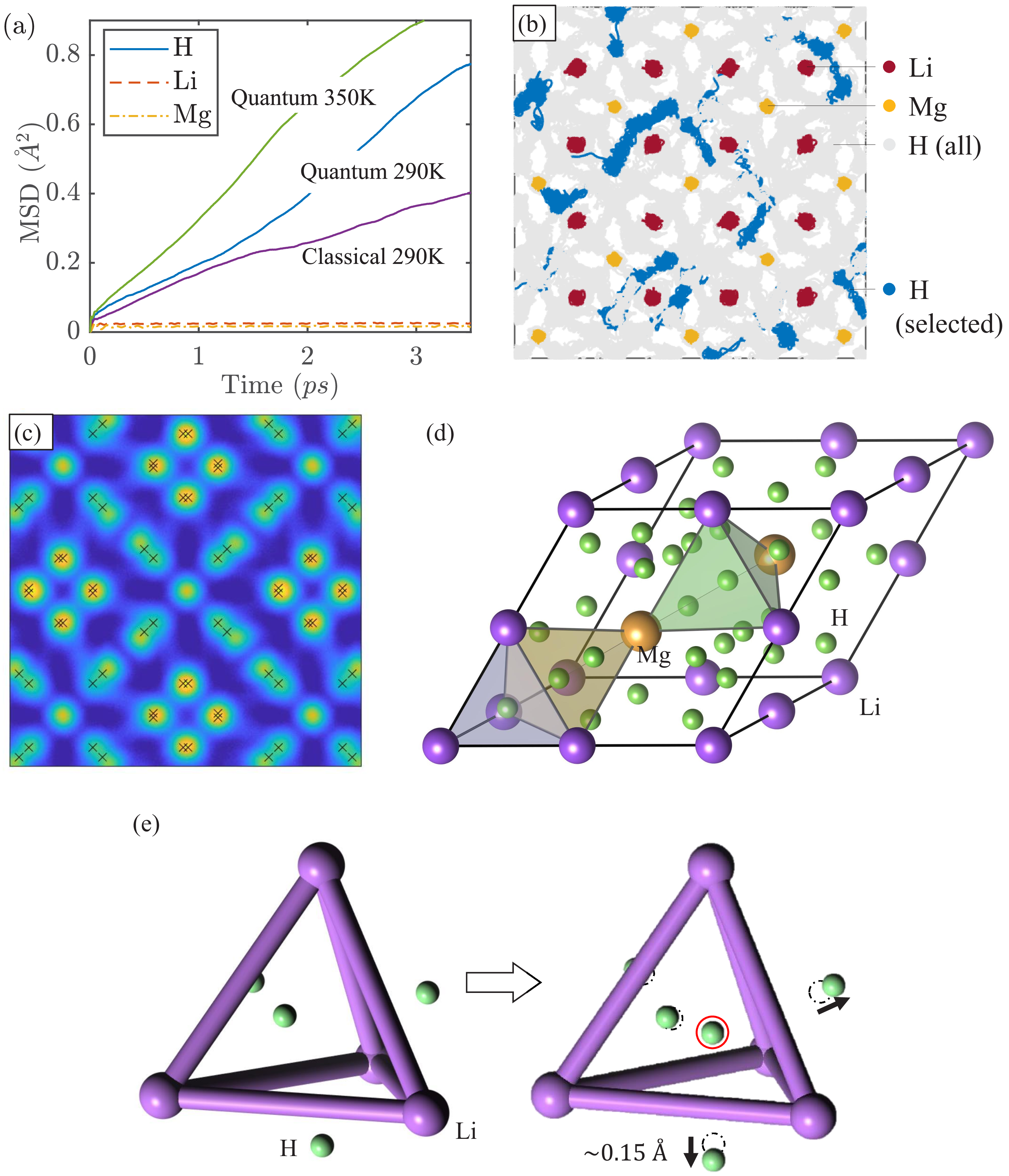}
	\caption{\label{fig:density} 
		(a) Mean squared displacement (MSD) of hydrogen (blue solid line), lithium (red dashed line) and magnesium (yellow dot-dashed line) ions under 260~GPa at 290~K. 
		MSD of hydrogen ions from classical MD simulation at 290~K and PIMD simulation at 350~K are also shown for comparison.
		(b) [100] view of trajectories of centroid mode of all hydrogen (grey), lithium (red) and magnesium (yellow) ions. Selected trajectories of hydrogen ions are marked by blue lines. 
		(c) [100] view of density distribution of hydrogen ions from a PIMD simulation. Hydrogen ions of the solid phase are marked with cross marks.
		(d) The effective `crystal structure' of the superionic phase. The structure is made up of three basic components: Li$_4$ (grey), Li$_3$Mg (yellow) and Li$_2$Mg$_2$ (green) tetrahedra, with hydrogen ions occupying their centers.
		(e) A schematically illustration of the effective structural transition around the Li$_4$ tetrahedra from solid phase (left) to superionic phase (right). In the right panel, the previously empty 8b site is circled in red solid line.
	}
\end{figure}

In the PIMD simulation, each quantum ion is mapped to a classical ring polymer~\cite{chandler_exploiting_1981}.
In Fig.~\ref{fig:density}(b), we show the trajectories of the center of mass of the ring polymers at 290~K, from which the mean squared displacements (MSD) of ions are determined.
It can be seen that all lithium and magnesium ions keep vibrating around their equilibrium positions, while hydrogen ions diffuse among the Li$_2$Mg lattice.
Correspondingly, the MSD of hydrogen ions keeps increasing, while those of lithium and magnesium ions remain around zero.
On average, the diffusion coefficient of hydrogen ions is $3.7\times10^{-6}$ $\mathrm{cm^2/s}$ at 290~K. 
Compared with the classical molecular dynamics result $1.9\times10^{-6}$ $\mathrm{cm^2/s}$, quantum effects accelerates the ion diffusion by about two times~\footnote{Both quantum and classical diffusion coefficients are obtained from simulations using (path-integral) Langevin thermostats, and are only used to schematically demonstrate the quantum effect of hydrogen ions. To correctly calculate the real-time dynamical properties, average over classical \textit{NVE} simulations and quantum techniques like ring-polymer molecular dynamics or centroid molecular dynamics are necessary~\cite{wang_quantum_2021,li_computer_2018}. When calculating $T_c$ using SPIA, only imaginary-time information is necessary and can be correctly captured using Langevin thermostats.}.

To further probe how the dynamical ions arrange in the system, we analyze the hydrogen density distributions. 
From Fig.~\ref{fig:density}(c), we see that density peaks form periodical structures, and are connected with each other by almost straight paths.
It appears that, instead of freely diffusing in the lattice like liquids, hydrogen ions tend to hop between sites.
By fitting the density distribution to three-dimensional Gaussian functions, we extract the effective `crystal structure' of the superionic phase, as shown in Fig.~\ref{fig:density}(d).
We find that the structure is almost identical to the solid phase, with hydrogen occupying 32e and 96g sites at the center of Li$_3$Mg and Li$_2$Mg$_2$ tetrahedra.
Now, the previously empty 8b sites at the center of Li$_4$ tetrahedra are also occupied.
This is consistent with Wang \textit{et al.}'s observations in Ref.~\cite{wang_quantum_2021}.
Due to the occupation, the hydrogen tetrahedron around it formed by ions at 32e sites is pushed slightly larger (see Fig.~\ref{fig:density}(e)). The site coordinates are listed in Table~S1~\cite{noauthor_see_nodate}. The new `crystal structure' still obeys the symmetry of $Fd\bar{3}m$ space group. 

\begin{figure}
	\includegraphics[width=8.6cm]{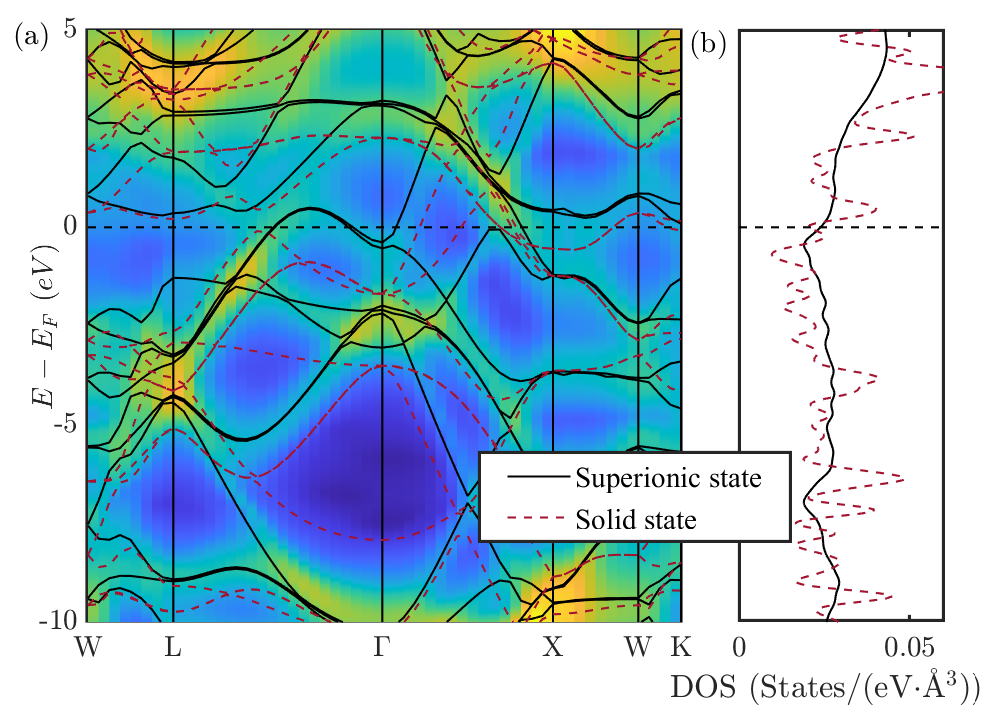}
	\caption{\label{fig:band} (a) Partial density of states $\rho(\bm{k},E)$ in the IBZ calculated under quasi-static approximation (mapped into colors in the background) and band structure solved from the effective Hamiltonian (black solid lines) in the superionic phase. Band structures in the solid phase (red dashed lines) is shown for comparison. (b) DOS in the two states. The superionic one is calculated using eigenvalues from all configurations.}
\end{figure}

\textit{Renormalized electronic structure.}—
We then study the electronic structure of the system. 
As analyzed above, while some ions diffuse between sites, the sites form a solid-like periodic framework.
As a result, system still has discrete translational symmetry, and electron states can be labeled by a wavevector $\bm{k}$ in the irreducible Brillouin zone (IBZ) of the primitive cell~\cite{allen_recovering_2013,zacharias_fully_2020}.
The band structure can be extracted from the partial density of states (DOS) $\rho(\bm{k},E)$ at $\bm{k}$:
\begin{eqnarray}
	\rho(\bm{k},E)=\frac{1}{2\pi} \mathrm{Tr}_{_{\bm{G}}} A(\bm{k}+\bm{G},\bm{k}+\bm{G}',E),
\end{eqnarray}
where $\hat{A}(E)$ is the electron spectral function, and $\bm{G}$($\bm{G}'$) is a reciprocal lattice vector of the primitive cell.
To see the qualitative properties of the band structure of Li$_2$MgH${}_{16}$,
we calculate $\rho(\bm{k},E)$ under the quasi-static approximation, i.e., instantaneous spectral functions are averaged over all ion configurations ~\cite{zacharias_fully_2020,noauthor_see_nodate}.
The result is shown in Fig.~\ref{fig:band}(a).
As expected, the band structure is strongly renormalized by ion diffusion and becomes qualitatively different from that in solid phase.
The full DOS is shown in Fig.~\ref{fig:band}(b).
We see that DOS at the Fermi surface is slightly lifted from 0.020~States$/eV/\AA^3$ to 0.024~States$/eV/\AA^3$ compared with the solid phase. Both values are high and are likely to support a large $\lambda$ and a high $T_c$.

On the other hand, due to the coupling with fluctuating ions, each band acquires a finite width, which are related to the inverse lifetime of quasi-electrons.
From the approximated $\rho(\bm{k},E)$, we find that the half width $\gamma$ is about 0.51~eV for states near the Fermi surface, while the Fermi energy $\varepsilon_F$ is about 23~eV.
Compared with the anharmonic solid H$_3$S, where $\gamma\sim$~0.16~eV and $\varepsilon_F\sim$~25~eV, the half width are larger, but still much smaller than the Fermi energy.
The observation indicates that long-lived quasi-electrons persist in the system, 
although the ion fluctuation in Li$_2$MgH${}_{16}$ is much stronger than ordinary solids.
This allows us to take a similar process of studying superconductivity as conventional theories. 
First, we study the propagation of quasi-electrons in the normal state. 
Then we determine the effective attractive interaction induced by the coupling with moving ions (phonon in the case of ordinary solids).

The non-perturbative SPIA provides a good framework to perform such analysis~\cite{liu_superconducting_2020,chen_first-principles_2021,chen_stochastic_2022,zhang_nonperturbative_2022}.
In SPIA, we calculate the Green's function $\bar{\mathcal{G}}(i\omega_j)$ to describe the propagation of quasi-electrons in normal states, where $\omega_j=(2j+1)\pi k_B T$ is a Fermion Matsubara frequency (See Supplementary Material~\cite{noauthor_see_nodate} for details).
Wave functions of the quasi-electrons are then naturally defined to approximately diagonalize $\bar{\mathcal{G}}(i\omega_j)$, 
so that the waves can propagate in the system without being scattered.
Generally, the Green's function has the structure $\bar{\mathcal{G}}(i\omega_j)=(i\omega_j-\hat{H}_{\mathrm{kin}}-\hat{\Sigma}(i\omega_j))^{-1}$, where $\hat{H}_{\mathrm{kin}}$ is the kinetic energy and $\hat{\Sigma}$ is the electron self energy. The Green's function is generally non-hermitian, and the eigenstates are thus not orthonormalized.
Fortunately, the small inverse lifetime $\gamma$ indicates a small anti-hermitian part of $\hat{H}_{\mathrm{kin}}+\hat{\Sigma}$. Therefore, we can diagonalize an effective Hamiltonian $\hat{H}_{\mathrm{eff}}=\hat{H}_{\mathrm{kin}}+\mathrm{Re}\hat{\Sigma}(i\omega_j)$
to determine the wave functions of quasi-electrons~\cite{chen_stochastic_2022}.
As the system is periodic, eigenstates of $\hat{H}_{\mathrm{eff}}$ take the form of Bloch waves.
The obtained band structure is shown in Fig.~\ref{fig:band}(a).
The dispersions coincide well with where the previously obtained $\rho(\bm{k},E)$ peaks. 

\textit{EPC parameters and superconductivity.}— 
We then study the superconductivity in the system.
We focus on the pairing between time reversal states $1\equiv(n,\bm{k},i\omega_j)$ and $\bar{1}\equiv(n,-\bm{k},-i\omega_j)$, where  $n$ and $\bm{k}$ are band and wavevector indices of the Bloch waves solved from $\hat{H}_{\mathrm{eff}}$.
The pairing is induced by the effective attractive interaction $\hat{W}$ mediated by moving ions.
In the framework of SPIA, $\hat{W}$ is determined by solving a Bethe-Salpeter equation,
\begin{eqnarray}\label{BSE}
	W_{11'}=\Gamma_{11'} + \frac{1}{\hbar^2\beta}\sum_2 W_{12}|\bar{\mathcal{G}}_2|^2\Gamma_{21'},
\end{eqnarray}
where $\Gamma_{11'}=-\beta\left\langle\mathcal{T}_{11'}\mathcal{T}_{\bar{1}\bar{1'}}\right\rangle$ is the fluctuation of $T$ matrices of electron-ion scattering in PIMD simulations~\cite{mahan_many-particle_2000}. It describes the scattering amplitude of a time-reversal quasi-electron pair from $(1,\bar{1})$ to $(1',\bar{1}')$~\cite{liu_superconducting_2020,chen_first-principles_2021,chen_stochastic_2022}.
The EPC parameters $\lambda(j-j')$ are then defined as the summation of $\hat{W}$ over the Fermi surface, i.e.,
\begin{multline}\label{lambda_avg}
	\lambda(j-j')=-\frac{1}{N(\varepsilon_F)}\sum_{n\bm{k},n'\bm{k'}}
	W_{n\bm{k},n'\bm{k'}}(j-j')\\
	\times\delta(\varepsilon_{n\bm{k}}-\varepsilon_F)\delta(\varepsilon_{n'\bm{k'}}-\varepsilon_F),
\end{multline}
where $j-j'$ represents a Bosonic Matsubara frequency $\nu_{j-j'}=\omega_{j}-\omega_{j'}$, and $N(\varepsilon_F)$ is the DOS on the Fermi surface. 
Similarly to conventional theories, the EPC parameters enter the linearized Eliashberg equations to determine $T_c$~\cite{allen_transition_1975,noauthor_see_nodate}.

\begin{figure}
	\includegraphics[width=8.6cm]{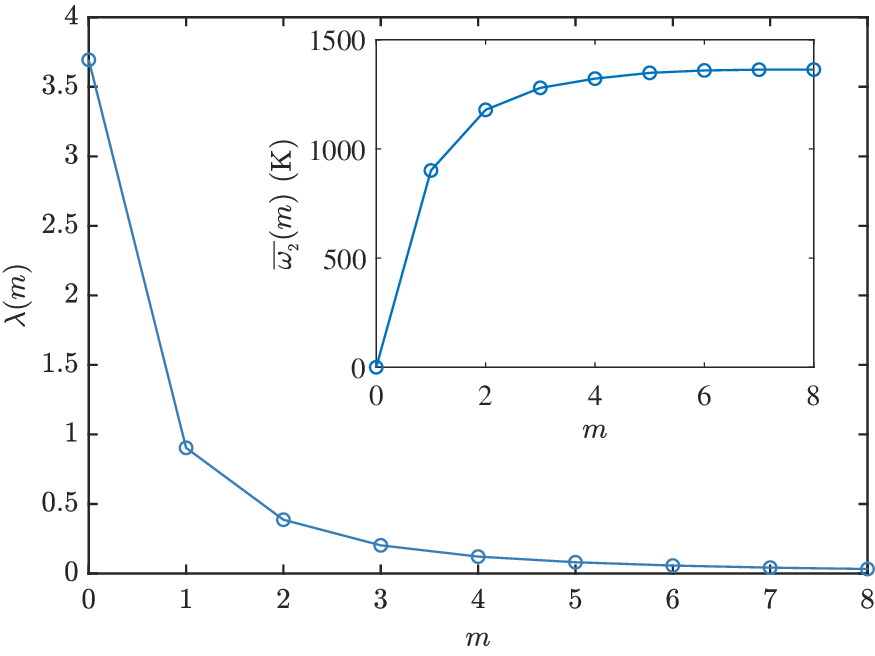}
	\caption{\label{fig:lam} EPC parameters $\lambda(m)$ of Li$_2$MgH${}_{16}$ under 260~GPa at 290~K, calculated on a $8\times8\times8$ k-grid and $8\times8\times8$ q-grid. Inset: the asymptotic behavior of $\bar{\omega}_2(m)=2\pi/\hbar\beta\sqrt{m^2\lambda(m)/\lambda(0)}$. The value at infinity $\bar{\omega}_2(\infty)$ gives the EPC-weighted average phonon frequency~\cite{allen_transition_1975,liu_superconducting_2020}.
		The results are calculated using non-diagonal supercells, each containing 304 atoms~\cite{noauthor_see_nodate}. }
\end{figure}

In practice, the Fermi surface summations in Eq.~(\ref{lambda_avg}) require dense sampling of wavevector transfer $\bm{q}=\bm{k}-\bm{k'}$.
The sampling is enabled by combining SPIA with MLFF~\cite{jinnouchi_--fly_2019}, which makes simulations in large supercells computationally affordable, and the non-diagonal supercell technique~\cite{lloyd-williams_lattice_2015}, which uses smaller supercells to cover a dense q-grid (see Supplementary Material~\cite{noauthor_see_nodate} for details).
By applying the approaches, we find that the static EPC parameter $\lambda(0)$=3.7. The value is close to that calculated in the solid phase $\lambda_{\mathrm{solid}}(0)\approx$4.0~\cite{sun_route_2019}, and much higher than the previous estimated value $\lambda(0)=1.6$ in superionic phase~\cite{wang_quantum_2021}. Values of higher-frequency components are shown in Fig.~\ref{fig:lam}.
Using a typical Morel-Anderson Coulomb pseudopotential $\mu^{*}=0.10$, we solve the linearized Eliashberg equations and find $T_c=277$~K~\footnote{In solving the Eliashberg equation, the Morel-Anderson pseudopotential is renormalized with respect to the cutoff $\omega_N=2\pi Nk_BT$ to be $1/\mu^{*}(N)=1/\mu^{*}+\mathrm{ln}(\overline{\omega_2}/\omega_N)$~\cite{allen_transition_1975}. This guarantees that $T_c$ does not rely on the energy cutoff. In our calculations, $\mu^{*}=0.16-0.10$ leads to $T_c=236-277$~K.}.
The value is much lower than the harmonic result $T_c=473$~K in the low-temperature solid phase~\cite{sun_route_2019}, but still close to the room-temperature regime.
At the temperature, Li$_2$MgH$_{16}$ is superionic, suggesting that the material is a superionic superconductor.

In the superionic metal, the behavior of ion motion is expected to  be temperature-dependent. 
For example, as shown in Fig.~\ref{fig:density}, the diffusion coefficient depends strongly on the temperature.
To examine the sensitivity of the predicted $T_c$ to the simulation temperature, we perform simulations at 350 K, which is 73 K higher than $T_c$. 
The diffusion coefficient is about 20\% higher than that at 290~K.
By assuming that the properties of ion motion do not change much at different temperatures, we represent EPC parameters $\lambda(m,T)$ as a single frequency-dependent function $\lambda\bm{(}\nu_m(T)\bm{)}$, and interpolate it to obtain values at other temperatures (see Fig. S1~\cite{noauthor_see_nodate}).
We find that the static EPC parameter $\lambda_{\mathrm{350K}}(0)$ is slightly suppressed to be $3.5$, while $T_c$ is overestimated by about 9~K (3~\% of $T_c$).

\begin{figure}
	\includegraphics[width=8.6cm]{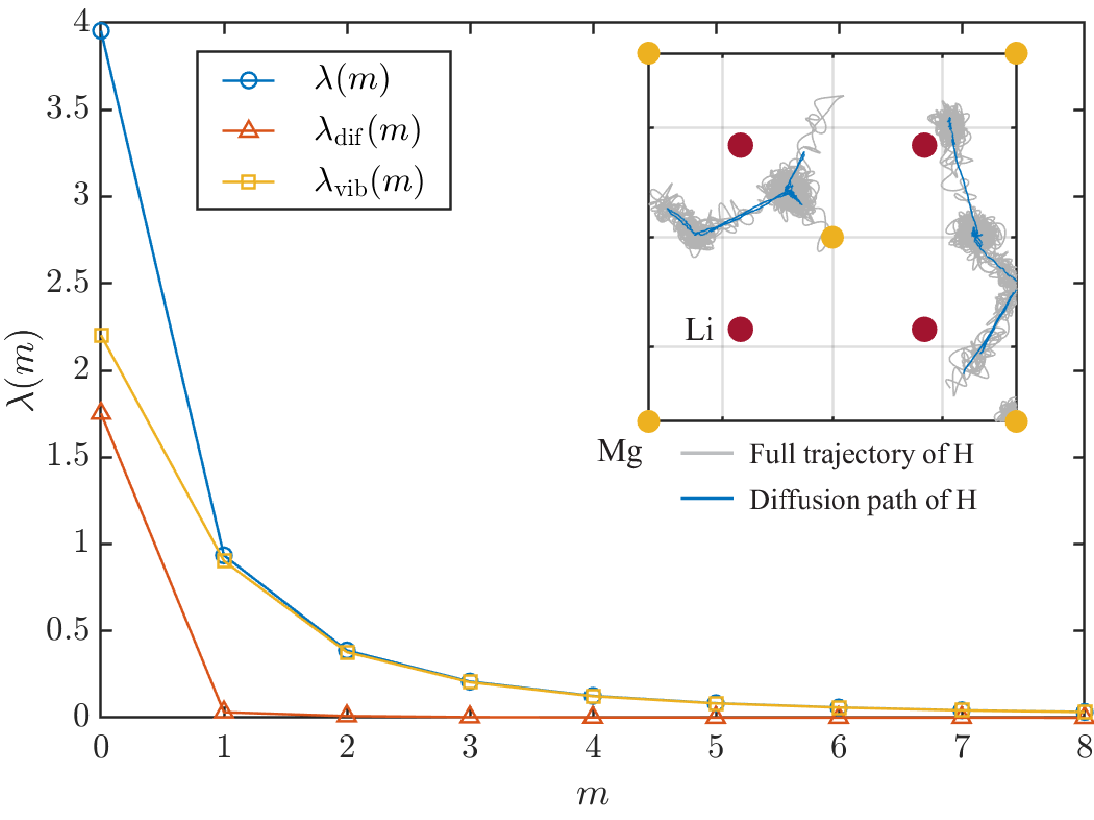}
	\caption{\label{fig:dif} 
		Contributions of different type of ion motion to the EPC parameters $\lambda(m)$ under 260~GPa at 290~K. Inset: full trajectories and diffusion paths of two selected hydrogen ions obtained using a moving-average filter with 250-fs window length. The diffusion contribution $\lambda_{\mathrm{dif}}(m)$ is calculated from all such diffusion paths. The results are calculated in a supercell containing 76 atoms.
	}
\end{figure}

To better understand the results, it is natural to ask how different types of ion motion contribute to superconductivity in Li$_2$MgH${}_{16}$. 
To answer the question, we perform further analysis in a smaller supercell containing 76 atoms. 
We assume that the ion trajectory can be expressed as the summation of diffusion paths and local vibrations around the path:
\begin{eqnarray}
	\bm{R}(t,\tau)=\bm{R}_{\mathrm{dif}}(t,\tau)+\Delta\bm{R}_{\mathrm{vib}}(t,\tau).
\end{eqnarray}
Since local vibrations are much faster than the diffusion,
the diffusion path can be obtained by performing a moving average of ion trajectories (see inset of Fig.~\ref{fig:dif}).
We then calculate the contribution of pure diffusion to EPC parameters $\lambda(m)$.
As shown in Fig.~\ref{fig:dif}, diffusion contribute a large $\lambda_{\mathrm{dif}}(0)=$1.75, but almost vanishing higher-frequency components.
Subtracting $\lambda_{\mathrm{dif}}(m)$ from $\lambda(m)$, we obtain contributions from local vibrations and vibration-diffusion couplings. The predicted $T_c$ is only slightly suppressed from 283~K to 275~K.
Based on the observations, we conclude that the main effect of ion diffusion is to renormalize electron wave functions and band structures.
The pairing between the electrons, however, is mainly induced by local vibrations. Diffusion greatly enhance $\lambda(0)$ but barely affect $T_c$. 
It indicates that a large $\lambda(0)$ is not a reliable predictor for $T_c$ for superionic solids.

\textit{Summary.}— In summary, we systematically study the ion motion, electronic structure and superconductivity in the superionic metal Li${}_2$MgH${}_{16}$ using first-principles calculations.
We find that, while the ions fluctuate much more strongly than conventional solids, long-lived quasi-electrons persist in the system. The wavefunctions and band structures of quasi-electrons are strongly renormalized.
The pairing between the quasi-electrons is analyzed using the non-perturbative SPIA method, so that all anharmonic effects like anharmonic local vibrations, quantum fluctuations and ion diffusion are taken into account properly.
By combining SPIA method with MLFF and non-diagonal supercell techniques, we are able to calculate electron-electron effective interactions on sufficiently dense k-grids and q-grids in such a complex system.
The superionic metal is predicted to have a large EPC parameter $\lambda(0)=3.7$ and a high superconducting $T_c=277$~K at 260~Gpa.
Further analysis suggests that ion diffusion contribute to the large $\lambda(0)$, while pairing between electrons is mainly induced by local vibrations of ions.
Our study predicts that Li$_2$MgH$_{16}$ is a superionic superconductor.
\nocite{li_computer_2018,liu_superconducting_2020,chen_first-principles_2021,chen_stochastic_2022,allen_transition_1975,migdal_interaction_1958,scalapino_superconductivity_1969,blochl_projector_1994,kresse_ultrasoft_1999,kresse_efficient_1996,perdew_generalized_1996,jinnouchi_--fly_2019,jinnouchi_phase_2019,ceriotti_efficient_2010,lloyd-williams_lattice_2015,sun_route_2019,allen_recovering_2013}

\begin{acknowledgments}
	We gratefully acknowledge Hanyu Liu, Jian Lv and Ying Sun for valuable
	discussions.
	The authors are supported by the National Science Foundation of China under Grant No.~12174005, the National Key R\&D Program of China under Grand Nos.~2018YFA0305603 and 2021YFA1401900. The computational resources were provided by the High-performance Computing Platform of Peking University.
\end{acknowledgments}

\bibliographystyle{apsrev4-2}
\bibliography{Reference}

\end{document}


	
	\title{Supplemental Materials:\\ Coexistence of Superconductivity and Superionicity in Li$_2$MgH$_{16}$}
	
	
	\author{Haoran Chen}
	\affiliation{International Center for Quantum Materials, Peking University, Beijing 100871, China}
	
	\author{Junren Shi}
	\email{junrenshi@pku.edu.cn}
	\affiliation{International Center for Quantum Materials, Peking University, Beijing 100871, China}
	\affiliation{Collaborative Innovation Center of Quantum Matter, Beijing 100871, China}
	
	
	
	\maketitle
	
	\setcounter{equation}{0}
	\setcounter{figure}{0}
	\setcounter{table}{0}
	\setcounter{page}{1}
	\makeatletter
	\renewcommand{\theequation}{S\arabic{equation}}
	\renewcommand{\thefigure}{S\arabic{figure}}
	\renewcommand{\thetable}{S\arabic{table}}
	\renewcommand{\bibnumfmt}[1]{[S#1]}
	\renewcommand{\citenumfont}[1]{S#1}
	\renewcommand{\thesection}{S\Roman{section}}
	
	\section{Stochastic path-integral approach}
	\subsection{Basic formulas}
	In this section, we outline the basic formulas of stochastic path-integral approach (SPIA). 
	Detailed derivations and discussions of the theory can be found in Refs.~\cite{liu_superconducting_2020,chen_first-principles_2021,chen_stochastic_2022}.
	
	First, we determine the electron Green's function $\hat{\mathcal{G}}$ at each ion configuration. Under the quasi-static approximation~\cite{liu_superconducting_2020,chen_first-principles_2021}, it is calculated as
	\begin{eqnarray}\label{Green_tau}
		\hat{\mathcal{G}}_j(\tau_i)
		=\hbar\left[(i\hbar\omega_j+\varepsilon_F)\hat{\mathbb{I}}-\hat{H}\big(\bm{R}(\tau_i)\big)\right]^{-1},
	\end{eqnarray}
	which is the instantaneous Green's function at time $\tau_i$. Here $\bm{R}(\tau_i)$ denotes the instantaneous ion positions at $\tau_i$, and $\hat{H}\big(\bm{R}(\tau_i)\big)$ is the corresponding Hamiltonian. $\omega_j=(2j+1)\pi/\hbar\beta$ is a large Fermionic Matsubara frequency satisfying $\hbar\omega_{ph}\ll|\hbar\omega_j|\ll\varepsilon_F$, where $\omega_{ph}$ is the characteristic phonon frequency of the system.
	
	The physical Green's function is then determined as the average of the instantaneous Green's functions over all ion configurations, i.e.,
	\begin{eqnarray}\label{Green}
		\hat{\bar{\mathcal{G}}}_j=\left\langle\frac{1}{N_b}\sum_{i=1}^{N_b}\hat{\mathcal{G}}_j(\tau_i)\right\rangle_C,
	\end{eqnarray}
	where $\langle\cdots\rangle_C$ denotes the configuration average.
	
	The physical Green's function $\hat{\bar{\mathcal{G}}}$ defines an effective medium in which electrons propagate. The scattering $T$ matrix with respect to the medium is determined by the relation
	\begin{eqnarray}\label{T_v2}
		\hat{\mathcal{T}}_j(\tau_i)
		=\hbar\hat{\bar{\mathcal{G}}}_j^{-1}
		\left(\hat{\mathcal{G}}_j(\tau_i)-\hat{\bar{\mathcal{G}}}_j\right)
		\hat{\bar{\mathcal{G}}}_j^{-1}.
	\end{eqnarray}
	Its value in the frequency domain is then obtained by performing a Fourier transformation, i.e.,
	\begin{eqnarray}\label{T_Fourier}
		\hat{\mathcal{T}}_{j'j}=
		\frac{1}{\hbar\beta}\int_0^{\hbar\beta}\hat{\mathcal{T}}_j(\tau)e^{i\nu_m\tau}d\tau,
	\end{eqnarray}
	with $\nu_m\equiv\omega_{j'}-\omega_{j}$.
	
	Assume that the Green's function is diagonalized by a set of generalized Bloch bases $|n\bm{k}\rangle$. 
	Under the bases, the pair scattering amplitude is 
	\begin{eqnarray}\label{Gamma_v2}
		\Gamma_{11'}
		=-\beta\left\langle|\mathcal{T}_{11'}|^2\right\rangle_C,
	\end{eqnarray}
	where $1\equiv(n\bm{k},i\omega_j)$.
	It describes the scattering of a time-reversal pair $(1,\bar{1})$ to $(1',\bar{1}')$.
	The effective interaction is then solved from a Bethe-Salpeter equation:
	\begin{eqnarray}\label{BSE_v2}
		W_{11'}=\Gamma_{11'} + \frac{1}{\hbar^2\beta}\sum_2 W_{12}|\bar{\mathcal{G}}_2|^2\Gamma_{21'}.
	\end{eqnarray}
	The equation is also solved under the quasi-static approximation~\cite{liu_superconducting_2020}.
	
	Finally, the EPC parameters are determined by summing $\hat{W}$ on the Fermi surface, i.e.,
	\begin{eqnarray}\label{lambda_avg_v2}
		\lambda(j-j')=-\frac{1}{N(\varepsilon_F)}\sum_{n\bm{k},n'\bm{k'}}
		W_{n\bm{k},n'\bm{k'}}(j-j')
		\delta(\varepsilon_{n\bm{k}}-\varepsilon_F)\delta(\varepsilon_{n'\bm{k'}}-\varepsilon_F),
	\end{eqnarray}
	where $N(\varepsilon_F)=\sum_{n\bm{k}}\delta(\varepsilon_{n\bm{k}}-\varepsilon_F)$ is the density of states (DOS) on the Fermi surface. The parameters enter linearized Eliashberg equations to determine $T_c$~\cite{allen_transition_1975,liu_superconducting_2020}.
	
	\subsection{Linearized Eliashberg equation}
	The Eliashberg equations read
	\begin{eqnarray}\label{Eliash}
		\rho\Delta_n &=& \sum_{n'}
		\left[
		\lambda(j-j') - \mu^\star - \frac{\hbar\beta}{\pi}
		|\tilde{\omega}(j)| \delta_{jj'}
		\right]
		\Delta_{j'},\quad\\
		\tilde{\omega}(j) &=& \frac{\pi}{\hbar\beta}
		\left(
		2j + 1 + \lambda(0) + 2\sum_{i=1}^j \lambda(i)
		\right),\quad j \ge 0.\quad
	\end{eqnarray}
	To determine $T_c$, one needs to find the temperature at which a non-negative eigenvalue $\rho$ of the Eliashberg equation~(\ref{Eliash}) first appears. 
	Ideally, the equations are self-consistently solved, and $\lambda(j-j',T)$ is determined from PIMD simulations at each temperature $T$.
	
	This is obviously too computationally expensive to afford.
	In practice, we assume that vibrational properties remain unchanged within a small temperature range, so that EPC parameters at different temperatures can be determined by a single frequency-dependent function $\lambda(m,T) = \lambda\bm{(}\nu_m(T)\bm{)}$.
	By calculating $\lambda(m,T_0)$ at a given temperature $T_0$, we can interpolate the values to obtain the frequency-dependent function $\lambda(\nu_m(T))$.
	
\begin{figure}
		\includegraphics[width=8.6cm]{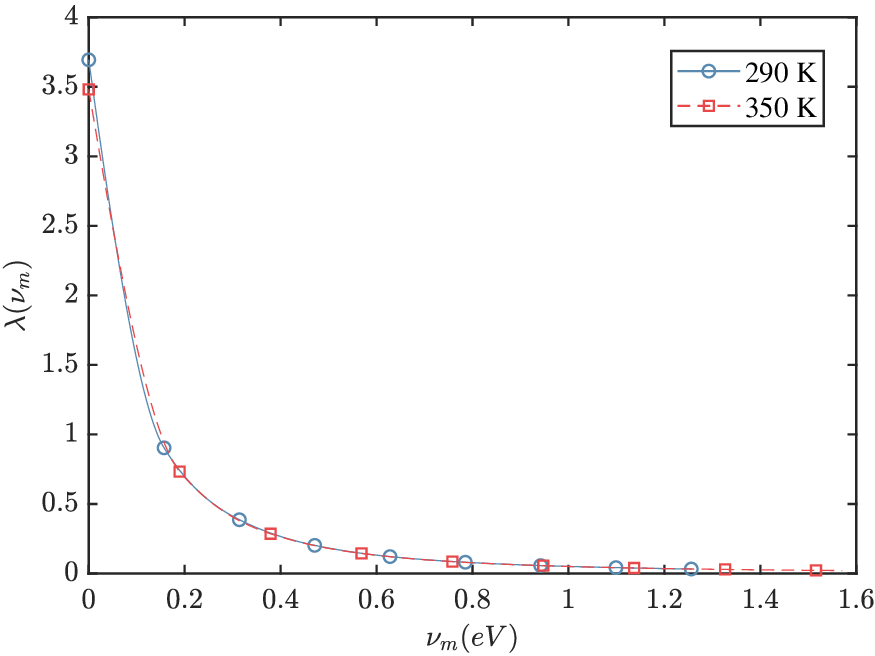}
		\caption{\label{fig:lam_T} EPC parameters $\lambda(\nu_m)$ from simulations at 290~K and 350~K, respectively.}
	\end{figure}

	In Fig.~\ref{fig:lam_T}, $\lambda(m,T_0)$ and the interpolated $\lambda_{T_0}(\nu_m(T))$ at $T_0=290$~K and $350$~K are shown. It can be seen that the two interpolated functions coincide well for $m>1$, while $\lambda_{\mathrm{350K}}(0)$ is slightly suppressed compared with $\lambda_{\mathrm{290K}}(0)$. Due to the temperature-dependent anharmonicity, $\lambda_{\mathrm{350K}}(\nu_m(T))$ also slightly overestimates $\lambda(1,T\approx T_c)$ and leads to an overestimation of $T_c$ by 9~K (3\% of $T_c$). 
	
	\subsection{Green's function in the PAW method}
	For balancing accuracy and efficiency, we implement our approach in the framework of PAW method~\cite{blochl_projector_1994,kresse_ultrasoft_1999}. 
	The PAW method establishes a connection between smooth pseudo (PS) wavefunctions $|\tilde{\psi}\rangle$ and the true all-electron (AE) waves $|\psi\rangle$ through a linear transformation operator:
	\begin{eqnarray}\label{Trans}
		|\psi\rangle=\hat{T}(\bm{R})|\tilde{\psi}\rangle,
	\end{eqnarray}
	where
	\begin{eqnarray}\label{Trans_op}
		\hat{T}(\bm{R})=\hat{\mathbb{I}}
		+\sum_{ia}\left(|\phi_{i}^a(\bm{R}_a)\rangle-|\tilde{\phi}_{i}^a(\bm{R}_a)\rangle\right)\langle \tilde{p}_i^a(\bm{R}_a)|\quad
	\end{eqnarray}  
	depends on the ion positions $\bm{R}\equiv\{\bm{R}_a\}$.
	Here $\langle\bm{r}|\phi_{i}^a(\bm{R}_a)\rangle=\phi_{i}^a(|\bm{r}-\bm{R}_a|)Y_{l_i m_i}(\widehat{\bm{r}-\bm{R}_a})$ is the AE partial wave with angular momentum $l_i m_i$ around ion $a$; $|\tilde{\phi}_{i}^a(\bm{R}_a)\rangle$ is the corresponding PS partial wave, and $|\tilde{p}_i^a(\bm{R}_a)\rangle$ is the projector function, which is constructed to be bi-orthonormal to PS partial waves. 
	With Eqs.~(\ref{Trans}, \ref{Trans_op}), formulas for quantities in the PS space can be constructed by transforming from the AE ones.
	
	Now we present the formula of Green's function in the framework of PAW method.
	By applying the transformation in Eq.~(\ref{Trans}),
	Eq.~(\ref{Green_tau}) can be transformed to a form that depends only on PS quantities~\cite{chen_first-principles_2021}, i.e.,
	\begin{eqnarray}\label{Green1}
		\hat{\mathcal{G}}(i\omega_j,\bm{R})=	\hat{T}(\bm{R})\hat{\tilde{\mathcal{G}}}(i\omega_j,\bm{R})\hat{T}^\dagger(\bm{R}),
	\end{eqnarray}
	where we define a PS Green's function:
	\begin{eqnarray}\label{Green2}
		\hat{\tilde{\mathcal{G}}}(i\omega_j,\bm{R})=\hbar\left[(i\hbar\omega_j+\varepsilon_F)\hat{S}(\bm{R})-\hat{\tilde{H}}(\bm{R})\right]^{-1},
	\end{eqnarray}
	with $\hat{\tilde{H}}(\bm{R})=\hat{T}^\dagger(\bm{R})\hat{H}(\bm{R})\hat{T}(\bm{R})$ being the PS Hamiltonian, and $\hat{S}(\bm{R})=\hat{T}^{\dagger}(\bm{R})\hat{T}(\bm{R})$ being the overlap matrix between PS waves.
	
	\subsection{Formula for superionic systems}\label{Formula_si}
	Equations~(\ref{Green1}) and~(\ref{Green2}) transform the Green's function to AE space, resulting in loss of high-energy information hidden in PS plane waves.
	To recover the information, we turn to a common PS space associated with the `equilibrium ion configuration' $\bm{R}^{(0)}$~\cite{chen_stochastic_2022}.
	In solids, this refers literally to the equilibrium positions of ions.
	In liquids, on the other hand, ion density distributes uniformly and all high-energy information vanishes on average. There is not an equilibrium configuration and the common PS space equals the AE space~\cite{chen_first-principles_2021}.
	In the intermediate superionic state, we define $\bm{R}^{(0)}\equiv\{\bm{R}^{\mathrm{solid-like}},\bm{R}^{\mathrm{density-peaks}}\}$, where $\bm{R}^{\mathrm{density-peaks}}$ refers to the positions of the density peaks of diffusing ions. 
	If the diffusing ions randomly walk around like in liquids, high-energy information vanishes likewise, and $\bm{R}^{\mathrm{density-peaks}}=$\,\O\ is an empty set.
	In Li$_2$MgH$_{16}$, $\bm{R}^{(0)}$ refers to the positions listed in Table~\ref{tab:crystal}.
	In the space, the matrix elements of Green's function become
	\begin{eqnarray}\label{Green_phi0}
		\left\langle\tilde{\psi}\left|
		\hat{S}(\bm{R}^{(0)},\bm{R})
		\hat{\tilde{\mathcal{G}}}(\mathrm{i}\omega_{j}, \bm{R})
		\hat{S}(\bm{R},\bm{R}^{(0)})
		\right|\tilde{\psi}^{\prime}\right\rangle,
	\end{eqnarray}
	where $\tilde\psi$ and $\tilde\psi^{\prime}$ are two states in the common PS space, and we define an overlap matrix:
	\begin{eqnarray}\label{overlap}
		\hat{S}(\bm{R},\bm{R}^{(0)}) = \hat{T}^\dagger(\bm{R})\hat{T}(\bm{R}^{(0)}).
	\end{eqnarray}
	The overlap matrix can be numerically calculated by substituting Eq.~(\ref{Trans_op}).
	Details of how to evaluate such quantities can be found in Sec.~III~A and Appendix~A of Ref.~\cite{chen_stochastic_2022}.
	
	\subsection{Reduction to conventional theory}
	Our approach can rigorously reduce to conventional formulas under harmonic and Migdal approximations. This can be seen by expanding the scattering ionic potential up to the first order of the displacement of ions, and apply our formulas. The derivation is as follows.
	
	Under harmonic approximations, the scattering ionic potential is expanded in terms of the displacement of ions $\hat{\mathcal{V}}=\hat{V}_{ei}(\bm{R})-\hat{V}_{ei}(\bm{R}^{\mathrm{eq}})\approx\Delta\bm{R}\cdot\partial{\hat{V}_{ei}}/\partial\bm{R}$. On the other hand, the displacements of ions can be expressed in terms of the phonon field operator $\hat{\phi}_{\bm{q}}$. The
	scattering potential can then be written as (see Eq. (32) of
	Ref.~\cite{giustino_electron-phonon_2017})
	\begin{eqnarray}\label{V11}
		\hat{\mathcal{V}}_{n\bm{k}j,n'\bm{k'}j'}= g_{nn'}(\bm{k'},\bm{q})\delta_{\bm{k},\bm{k'+q}}
		\frac{1}{\hbar\beta}\int_{0}^{\hbar\beta}d\tau \hat{\phi}_{\bm{q}}(\tau) \mathrm{e}^{-i(\omega_{j}-\omega_{j'})\tau},
	\end{eqnarray}
	where $\hat{\phi}_{\bm{q}}(\tau)$ is the phonon field operator, $\bm{k}(\bm{q})$ is the electron (phonon) momentum, $g$ is the electron-phonon coupling matrix, and $\omega_{j(j')}$ is a Fermionic Matsubara frequency. The effective electron-electron interaction in conventional harmonic theories is exactly the fluctuation of the scattering ionic potential:
	\begin{eqnarray}\label{W_har}
		W_{11'}^{\mathrm{har}}&&
		=-\beta\langle\mathcal{V}_{11'}\mathcal{V}_{\bar{1}\bar{1}'}\rangle_C\nonumber\\
		&&\approx|g_{nn'}(\bm{k'},\bm{q})|^2 D(\bm{q},\omega_j-\omega_{j'})\delta_{\bm{k},\bm{k'+q}},
	\end{eqnarray}
	where $D$ is the phonon Green’s function. Here we exploit the that a PIMD configuration average is equivalent to an imaginary-time ordered ensemble average~\cite{mahan_many-particle_2000}, i.e., $\langle\cdots\rangle_C=\langle\hat{T}_{\tau}[\cdots]\rangle$.
	The proof can be found in Sec. II B 2 of Ref.~\cite{liu_superconducting_2020}.
	
	Compare the formula with SPIA formulas:
	\begin{eqnarray}\label{W11_SPIA}
		W_{11'}=\Gamma_{11'}+\frac{1}{\hbar^2\beta}\sum_{2}W_{12}|\bar{\mathcal{G}}_2|^2\Gamma_{21'},
	\end{eqnarray}
	where
	\begin{eqnarray}
		\Gamma_{11'}=-\beta\langle\mathcal{T}_{11'}\mathcal{T}_{\bar{1}\bar{1}'}\rangle_C
	\end{eqnarray}
	and 
	\begin{eqnarray}
		\mathcal{T}_{11'}=\mathcal{V}_{11'}+
		\frac{1}{\hbar\beta}\sum_{2}\mathcal{V}_{12}\bar{\mathcal{G}}_2\mathcal{T}_{21'}.
	\end{eqnarray}
	We see that there are then two ways to build a bridge between with SPIA formulas. The first is rather straight forward. If we perform two Born approximations, i.e, $\mathcal{T}_{11'}\approx\mathcal{V}_{11'}$ and $W_{11'}\approx\Gamma_{11'}$, then Eq.~(\ref{W_har}) is recovered. This is the argument proposed in Sec. II B 5 of Ref.~\cite{liu_superconducting_2020}.
	
\begin{figure}
		\includegraphics[width=15.2cm]{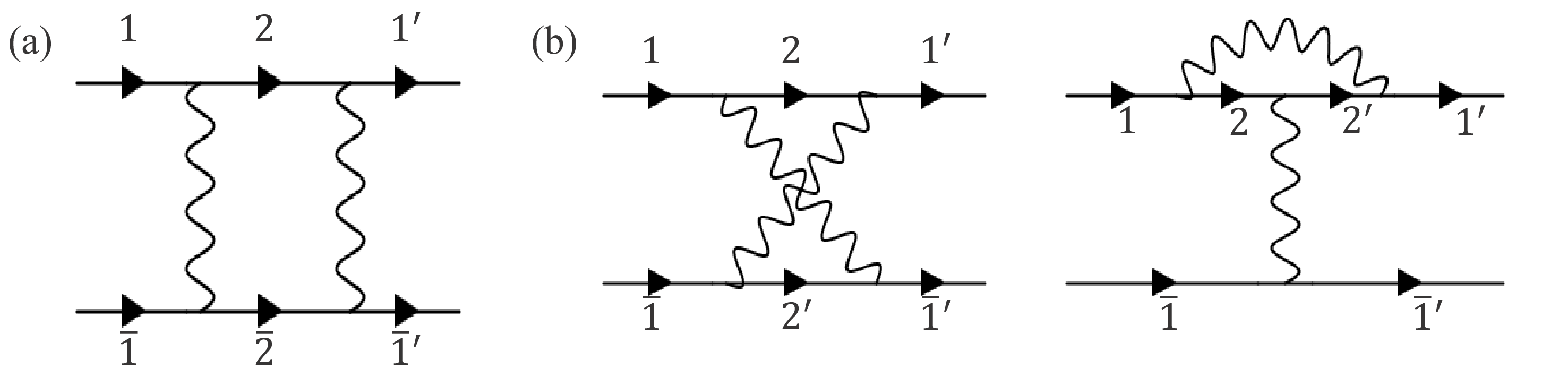}
		\caption{\label{fig:feynman} All second-order Feynman diagrams of $\Gamma_{11'}$ under harmonic approximations, which are derived from $\langle\mathcal{V}_{12}\bar{\mathcal{G}}_{2}\mathcal{V}_{21'}\mathcal{V}_{\bar{1}2'}\bar{\mathcal{G}}_{2'}\mathcal{V}_{2'\bar{1}'}\rangle_C$ and $\langle\mathcal{V}_{12}\bar{\mathcal{G}}_{2}\mathcal{V}_{22'}\bar{\mathcal{G}}_{2'}\mathcal{V}_{2'1'}\mathcal{V}_{\bar{1}\bar{1}'}\rangle_C$: (a) single-phonon processes corresponding to the first line of Eq.~(\ref{Gamma_har}); (b) multi-phonon processes, which are of the order $\sqrt{m_e/M_{\mathrm{ion}}}$ according to the Migdal theorem.}
	\end{figure}

	A more rigorous way is to directly substitute Eq.~(\ref{V11}) into the SPIA formulas, and solve the equations. Since we are dealing with a harmonic system, we can use Wick's theorem to evaluate the high-order correlations of $\mathcal{V}$. Finally, we get
	\begin{eqnarray}\label{Gamma_har}
		\Gamma_{11'}=&&W_{11'}^{\mathrm{har}}
		-\frac{1}{\hbar^2\beta}\sum_{2}W_{12}^{\mathrm{har}}\bar{\mathcal{G}}_2 W_{21'}^{\mathrm{har}}
		+\frac{1}{\hbar^4\beta^2}\sum_{22'}W_{12}^{\mathrm{har}}\bar{\mathcal{G}}_2 W_{22'}^{\mathrm{har}}\bar{\mathcal{G}}_{2'} W_{2'1'}^{\mathrm{har}} + \cdots \nonumber\\
		&& + \text{multi-phonon terms}.
	\end{eqnarray}
	The second line refers to terms involving multi-phonon scattering processes, like $\bm{(}g_{12}g_{2'\bar{1}}D(k_2-k_1)\bm{)}
	\bm{(}g_{21'}g_{\bar{1}2'}D(k_{1'}-k_2)\bm{)}$.
	If we represent the terms as Feynman diagrams, as shown in Fig.~\ref{fig:feynman}, the multi-phonon terms are those have crossing phonon lines. According to the Migdal theorem, these terms are of the order $\sqrt{m_e/M_{\mathrm{ion}}}$~\cite{migdal_interaction_1958,scalapino_superconductivity_1969}. They are called vertex corrections in conventional theories, and can be safely ignored in most cases. Compare the first line of Eq.~(\ref{Gamma_har}) with Eq.~(\ref{W11_SPIA}), it can be immediately seen that 
	\begin{eqnarray}
		W_{n\bm{k}j,n'\bm{k}'j'}=W_{n\bm{k}j,n'\bm{k}'j'}^{\mathrm{har}}=|g_{nn'}(\bm{k'},\bm{q})|^2 D(\bm{q},\omega_j-\omega_{j'})\delta_{\bm{k},\bm{k'+q}}.
	\end{eqnarray}
	
	\section{Numerical details}
	\subsection{Density functional theory calculations}
	All DFT calculations are performed using the CPU and GPU version of the Vienna ab initio Simulation Package (VASP) code~\cite{kresse_efficient_1996,kresse_ultrasoft_1999}. The projector-augmented wave (PAW) method~\cite{blochl_projector_1994,kresse_ultrasoft_1999} is used to describe the ion-electron interaction, and the Perdew-Burke-Ernzerhof (PBE) functional~\cite{perdew_generalized_1996} is used to describe the exchange-correlation effect.
	
	\subsection{Machine learning force field}
	We use the built-in module of VASP to train the force field on the fly~\cite{jinnouchi_--fly_2019,jinnouchi_phase_2019}. The program is modified to allow on-the-fly training in a path-integral molecular dynamics (PIMD) simulation.
	We construct a Born-von Karman $2\times2\times2$ supercell of Li${}_2$MgH${}_{16}$ and perform MD and PIMD simulations. 
	(Path-integral) Langevin thermostat is used to control temperature in simulations~\cite{ceriotti_efficient_2010}. The friction coefficients of centroid mode are set to [1 1 1]~ps${}^{-1}$ and a time step of 1~fs is used.
	During the simulation, if a structure is judged to be outside the previous training sets, a DFT calculation is performed.
	In DFT calculations, we use a large energy cutoff of 800~eV for plane waves to expand electron wavefunctions, and a dense $4\times4\times4$ $\Gamma$-centered $k$-point grid to sample the Brillouin zone of the supercell.
	
	The training is performed in three steps.
	First, an on-the-fly MD training is performed. The temperature is raised from 290~K to 450~K in a 20-ps simulation. Second, quantum effects are taken into account by performing a PIMD simulation with bead number (Trotter number) $N_b=8$. A short simulation of 2~ps at 220~K is first taken starting from the solid phase. Then another 20-ps simulation in the superionic phase is performed while raising the temperature from 350~K to 450~K. In these steps, the radial and angular descriptors are constructed using parameters ML\_SION=0.5 and ML\_MBR=8. In all, 666 structures and 253 (190,195) local configurations for Li (Mg,H) are chosen for training, and the root-mean-square error (RMSE) of forces reaches 0.203~eV$/\AA$. 
	Finally, we re-choose local configurations using ML\_SION=0.3 and ML\_MBR=12 to improve the accuracy of the force field. In this step, 1087 (633,4131) local configurations for Li (Mg,H) are chosen, and the RMSE is reduced to 0.121~eV$/\AA$. The force field is applied in our subsequent studies.
	
\begin{figure}
		\includegraphics[width=17.2cm]{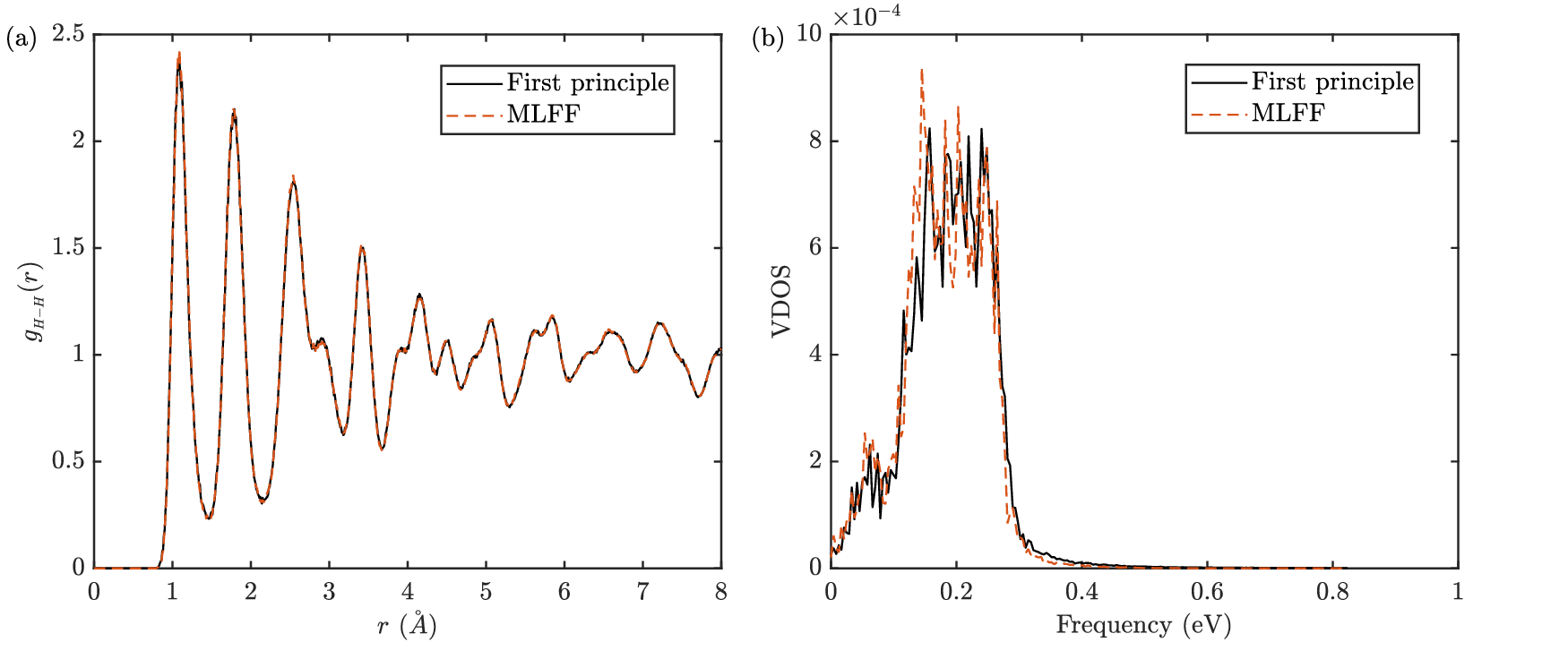}
		\caption{\label{fig:RDF} (a) Hydrogen-Hydrogen radial distribution function and (b) vibrational density of states (VDOS) of the centroid mode from DFT-based and MLFF-based PIMD, respectively.}
	\end{figure}

\begin{figure}
		\includegraphics[width=8.6cm]{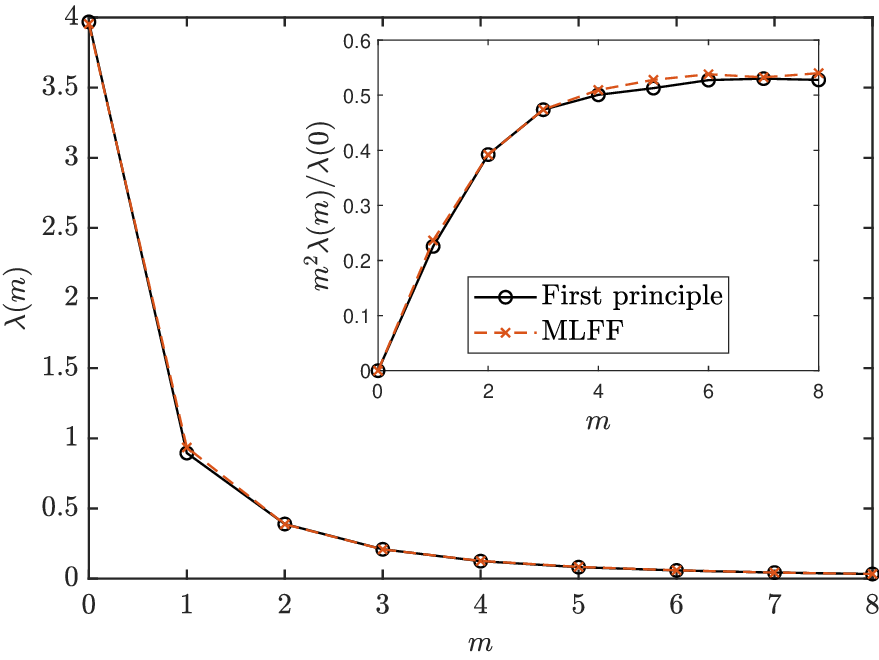}
		\caption{\label{fig:lam} EPC parameters $\lambda(m)$ from DFT-based and MLFF-based PIMD, respectively. The asymptotic behavior of $m^2\lambda(m)$ is also shown in the inset for comparison.}
	\end{figure}

	To test the force field, we perform two independent PIMD simulations in a non-diagonal supercell with $\hat{\mathbb{S}}=\{2,0,-1;0,1,0;0,0,1\}$ using DFT and MLFF, respectively. The simulations are performed at 290~K with bead number $N_b=16$.
	In Figs.~\ref{fig:RDF} and~\ref{fig:lam}, we compare radial distribution function (RDF), vibrational density of states (VDOS) and EPC parameters obtained the two simulations, respectively. Satisfactory agreements are obtained in all cases.
	
	\subsection{Effective interaction and EPC parameters}
	To apply SPIA method to Li$_2$MgH${}_{16}$, we first use the MLFF above to perform PIMD simulations. 17 non-diagonal supercells of different shapes are used to reach all irreducible $\bm{q}$ points of a $8\times8\times8$ grid~\cite{lloyd-williams_lattice_2015}. For each supercell, a time step of 0.5~fs and an overall simulation time of 3~ps with bead number $N_b$=16 is used to simulate the quantum system at 290~K and 350~K.
	
	The ion configurations are uniformly sampled with a spacing of 40 time steps after a 0.5-ps equilibration. DFT calculations are performed for these configurations. An energy cutoff of 500~eV for plane waves to expand electron wavefunctions is used. The density of $bm{k}$-point grids is chosen to reach an accuracy within 1~meV/atom. 
	The converged local densities are then used as inputs of our MATLAB implementation of SPIA.
	The effective electron-electron interaction $\hat{W}$ are then calculated on the irreducible wedge of a $8\times8\times8$ $\bm{k}$-point grid. 
	The Fermi-surface-averaged EPC parameters are then determined by summing $\hat{W}$ on the Fermi surface~\cite{chen_stochastic_2022}, i.e.,
	\begin{eqnarray}\label{lambda_avg}
		\lambda(j-j')=-\frac{1}{N(\varepsilon_F)}\sum_{n\bm{k},n'\bm{k'}}
		W_{n\bm{k},n'\bm{k'}}(j-j')
		\delta(\varepsilon_{n\bm{k}}-\varepsilon_F)\delta(\varepsilon_{n'\bm{k'}}-\varepsilon_F),
	\end{eqnarray}
	where the index $j$ denotes the Fermion Matsubara frequency $\nu_j=(2j+1)\pi/\beta$, with $\beta=k_B T$.
	The delta function $\delta(\varepsilon_{n\bm{k}}-\varepsilon_F)$ is replaced with a Lorentzian function
	\begin{eqnarray}
		\delta(\varepsilon_{n\bm{k}}-\varepsilon_F)\rightarrow \frac{1}{\pi}\frac{\gamma}{(\varepsilon_{n\bm{k}}-\varepsilon_F)^2+\gamma^2}
	\end{eqnarray}
	in our calculation. The half width is set to $\gamma=0.8$~eV to ensure that the calculated DOS matches that obtained from the eigenvalue distribution in the PIMD simulation~(Fig.~2(b) in the main text).

	We note that the half width is larger than that of regular solids, because the diffusing behavior of ions smears out the electron Fermi surface. If we reduce the half width from 0.8~eV to the usual values 0.2~-~0.4~eV in solids, $T_c$ will be reduced from 277~K to 250~-~263~K, when $\mu^{*}=0.10$.

	\section{Crystal structures}
	As mentioned in the main text, we fit the density distribution to three-dimensional Gaussian functions and extract the effective `crystal structure' of the superionic phase. The ion positions are listed in Table.~\ref{tab:crystal}. For comparison, we also list the solid crystal structure from Ref.~\cite{sun_route_2019}.
	
	\begin{table}
		\caption{\label{tab:crystal}%
			{Crystal structure of the solid phase of Li$_2$Mg${}_{16}$, and the effective one of the superionic phase.}}
		\begin{ruledtabular}
			\begin{tabular}{lcccccc}
				\multirow{2}{*}{\textbf{Phase}} & \multirow{2}{*}{\textbf{Space group}}	& 
				\multirow{2}{*}{\shortstack{\textbf{Lattice parameters}\\ \textbf{($\AA$, ${}^{^\circ}$)}}}     &
				\multicolumn{4}{c}{\textbf{Atomic coordinates (fractional)}}   \\
				& & & \textbf{Atoms} & \textbf{X} & \textbf{Y} & \textbf{Z}\\  
				\midrule
				\textbf{Solid phase} & Fd$\bar{3}$m & $a=b=c=$6.718513 & Li(16c) & 0.62500 & 0.87500  & 0.87500 \\
				\textbf{(250~GPa)} & & $\alpha=\beta=\gamma=$90.00 & Mg(8b) & 0.00000 & 0.00000 & 0.00000 \\
				& & & H(32e) & 0.83306 & 0.16694 & 0.16694 \\
				& & & H(96g) & 0.06466 & 0.24560 & 0.43534 \\
				\midrule
				\textbf{Superionic phase} & Fd$\bar{3}$m & $a=b=c=$6.718513 & Li(16c) & 0.62500 & 0.87500  & 0.87500 \\
				\textbf{(260~GPa)} & & $\alpha=\beta=\gamma=$90.00 & Mg(8b) & 0.00000 & 0.00000 & 0.00000 \\
				& & & H(8b) & 0.25000 & 0.25000 & 0.75000 \\
				& & & H(32e)& 0.84609 & 0.15391 & 0.15391 \\
				& & & H(96g) & 0.06318 & 0.24933 & 0.43682 \\
			\end{tabular}
		\end{ruledtabular}
	\end{table}
	
	\section{Partial density of states}
	The partial density of states (PDOS) $\rho(\bm{k},E)$ shown in Fig.~2~(a) in the main text is defined as~\cite{allen_recovering_2013}
	\begin{eqnarray}\label{rho_k}
		\rho(\bm{k},E) = \frac{1}{2\pi}\mathrm{Tr}_{{}_{\bm{G}}} A(\bm{k}+\bm{G},\bm{k}+\bm{G'},E),
	\end{eqnarray}
	where $A(\bm{k}+\bm{G},\bm{k}+\bm{G'},E)$ is the electron spectral function, and $\bm{G},\bm{G'}$ are reciprocal lattice vectors of the primitive cell. 
	An accurate determination of $\rho(\bm{k},E)$ is beyond the focus of this study. 
	To take a look at the qualitative behaviors, we simply use the quasi-static approximation and average all instantaneous static PDOS~\cite{liu_superconducting_2020}. 
	As a result, we note that the resulting $\rho(\bm{k},E)$ cannot be regarded as the real partial spectral function. 
	For example, the strong dependence of electronic self energy on $E$ within the scale of phonon energy~\cite{mahan_many-particle_2000} cannot be reproduced in our current approximate calculations.
	
	Now $\rho(\bm{k},E)$ is approximately determined as
	\begin{eqnarray}
		\rho(\bm{k},E)\approx
		\left\langle \rho\left(\bm{k},E;\{\bm{R}\}\right)\right\rangle_C,
	\end{eqnarray}
	where $\langle\cdots\rangle_C$ denotes an average over all configurations.
	$\rho\left(\bm{k},E;\{\bm{R}\}\right)$ is the corresponding instantaneous density of states
	\begin{eqnarray}\label{rhoR}
		\rho(\bm{k},E;\{\bm{R}\})=
		\sum_{N\bm{K},\bm{G}}
		\left|\left\langle\bm{k+G}\Big|\psi_{N\bm{K}}^{\{\bm{R}\}}\right\rangle\right|^2
		\delta\left(E-\varepsilon_{N\bm{K}}^{\{\bm{R}\}}\right),
	\end{eqnarray}
	where $\psi_{N\bm{K}}^{\{\bm{R}\}}$ is the instantaneous eigenstate labeled by band index $N$ and wavevector $\bm{K}$ of the supercell, with $\varepsilon_{N\bm{K}}^{\{\bm{R}\}}$ being the corresponding eigenvalue. If the supercell is made up of $M$ primitive cells, $\psi_{N\bm{K}}$ contains information of $M$ $\bm{k}$ points in the IBZ of the primitive cell. 
	
	In the framework of PAW method, the matrix becomes
	\begin{eqnarray}
		\left\langle\bm{k+G}\Big|\psi_{N\bm{K}}^{\{\bm{R}\}}\right\rangle
		=\left\langle\bm{k}+\tilde{\bm{G}}(\bm{R}^{(0)})\Big|
		\hat{T}^{\dagger}(\bm{R}^{(0)})\hat{T}(\bm{R})
		\Big|\tilde{\psi}_{N\bm{K}}^{\{\bm{R}\}}\right\rangle,
	\end{eqnarray}
	where $\bm{R}^{(0)}$ is defined in Sec.~\ref{Formula_si} and corresponds to the ion configuration listed in Table~\ref{tab:crystal}. $\hat{T}$ is a linear transformation operator that relates PS wavefunctions and AE wavefunctions, which is central in the PAW method~\cite{kresse_ultrasoft_1999}. Calculations of the matrix elements of $\hat{T}^{\dagger}(\bm{R}^{(0)})\hat{T}(\bm{R})$ can be found in Ref.~\cite{chen_stochastic_2022}.
	 
	In practice, the delta function is approximated by a Lorentzian function in our calculations. The half width is set to $\gamma=0.02$~Ry to achieve a balance between accuracy and computational cost.
	We note that this shifts the target energy from $E$ to $E+i\gamma$.

	\section{Tests of convergence}
	In this section, we test the convergence of our calculations with respect to the sampling time step, simulation time, bead number and k-mesh density.
	
\begin{figure}
		\includegraphics[width=17.2cm]{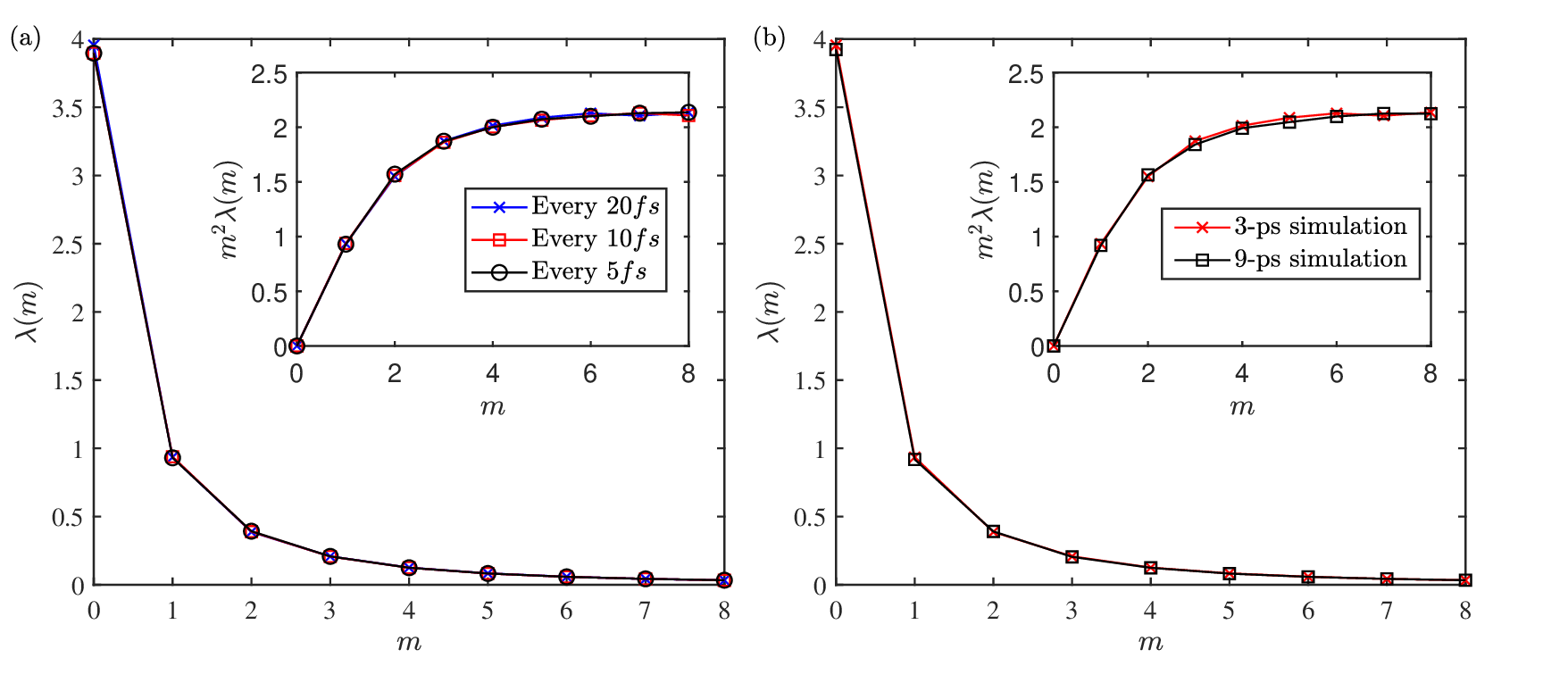}
		\caption{\label{fig:conv_time} EPC parameters $\lambda(m)$ with respect to (a) sampling time steps and (b) simulation time lengths. The asymptotic behavior of $m^2\lambda(m)$ is also shown in the inset for comparison.}
	\end{figure}

	First, we test the sampling time step when calculating Green's functions and effective interactions. Tests are performed in a non-diagonal supercell with $\hat{\mathbb{S}}=\{2,0,-1;0,1,0;0,0,1\}$ at 290~K.
	In Fig.~\ref{fig:conv_time}~(a), we show EPC parameters $\lambda(m)$ calculated with configurations sampled every 5~fs, 10~fs and 20~fs. We see that results from three calculations show little difference. The predicted $T_c$'s are 284~K, 284~K and 285~K, respectively. We thus use a sampling time step of 20~fs in all calculations.
	
	Second, we perform convergence tests regarding the simulation time length in the non-diagonal supercell above. In Fig.~\ref{fig:conv_time}~(b), we see that a simulation length of 3 ps is well converged.
	
\begin{figure}
		\includegraphics[width=17.2cm]{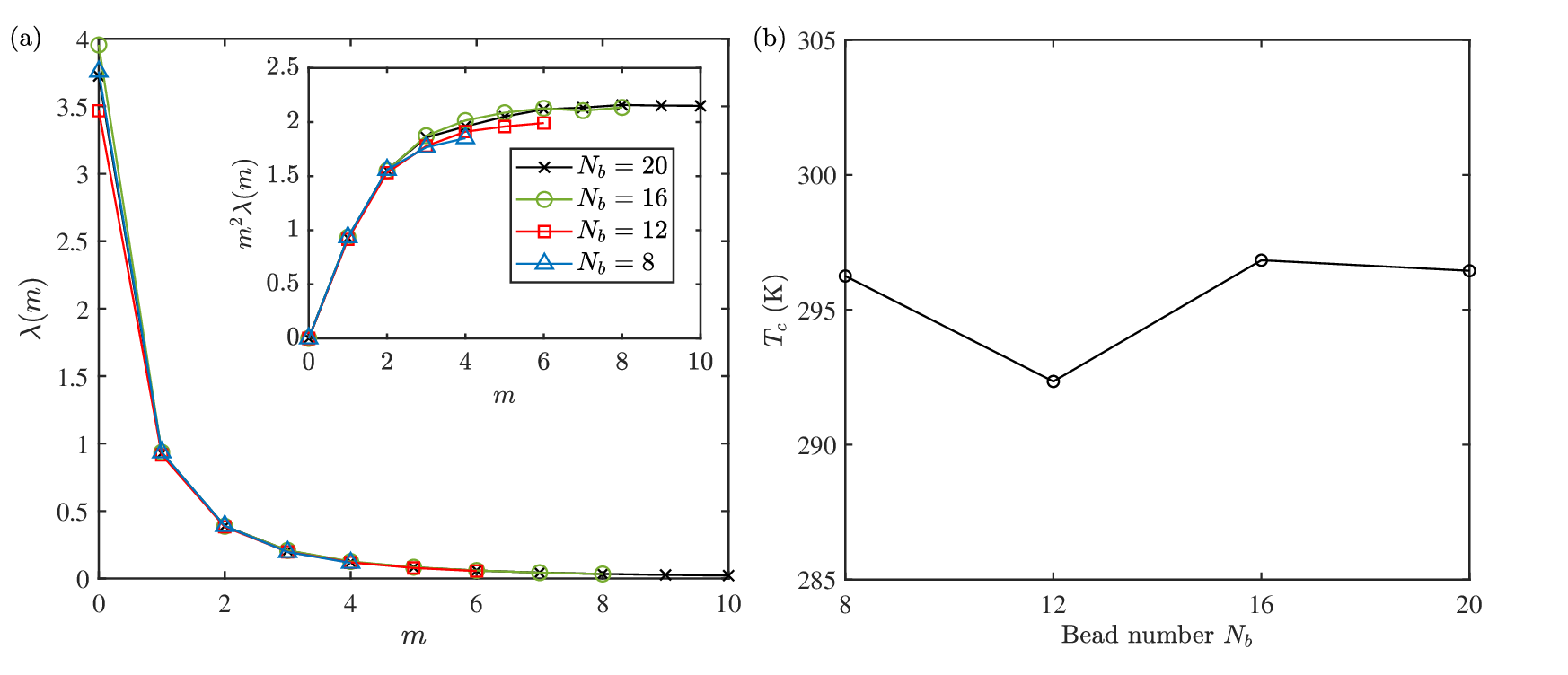}
		\caption{\label{fig:conv_bead} (a) EPC parameters $\lambda(m)$ with respect to bead number. The asymptotic behavior of $m^2\lambda(m)$ is also shown in the inset for comparison. (b) Superconducting $T_c$ with respect to bead number.}
	\end{figure}

	Third, we test the dependence of $\lambda$ and $T_c$ on bead number.  As shown in Fig.~\ref{fig:conv_bead}, both $\lambda$ and $T_c$ are converged at $N_b=16$.
	
	\begin{figure}
		\includegraphics[width=8.6cm]{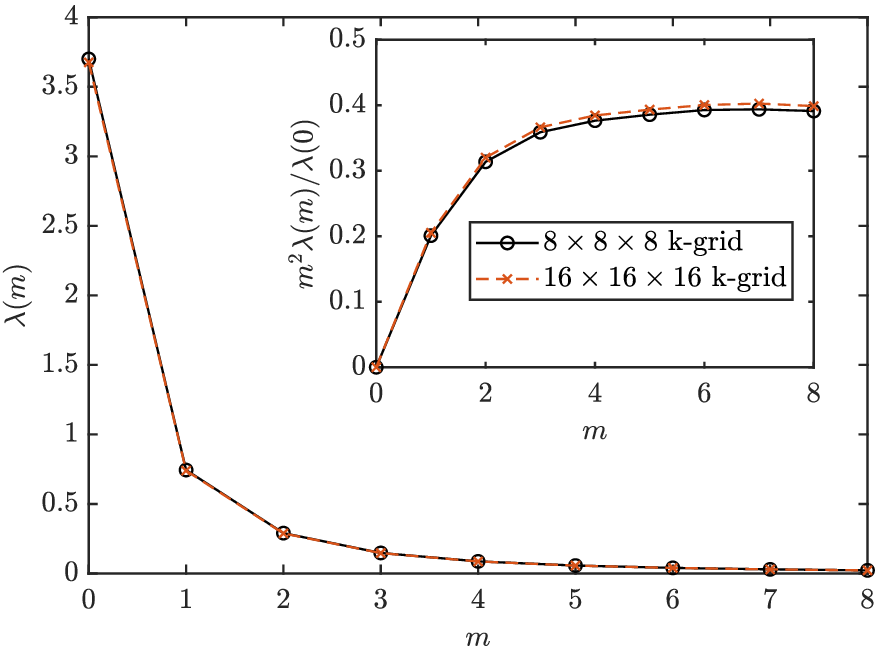}
		\caption{\label{fig:conv_k} EPC parameters $\lambda(m)$ calculated from a $8\times8\times8$ and $16\times16\times16$ k-grid, respectively. The asymptotic behavior of $m^2\lambda(m)$ is also shown in the inset for comparison.}
	\end{figure}
	
	Finally, we test whether the k-mesh density is sufficient for determining $\lambda$ and $T_c$. We use a non-diagonal supercell with $\mathbb{S}=\{1,0,0; 0,1,0; -2,-3,8\}$, which samples several $\bm{q}$ points on the $8\times8\times8$ mesh. The EPC parameters are then calculated on a $8\times8\times8$ and $16\times16\times16$ $\bm{k}$-point mesh of the primitive cell, respectively. As shown in Fig.~\ref{fig:conv_k}, the results from two calculations are almost the same.
	
	\bibliographystyle{apsrev4-2}
	\bibliography{Reference}